\documentclass{aa}  

\usepackage{graphicx}
\usepackage{txfonts}
\begin{document}

   \title{Water masers in Compton-thick AGN}

   \subtitle{I. Detailed study of the new water megamaser in IRAS~15480-0344}

   \author{P. Castangia
          \inst{1}
          \and
          A. Tarchi\inst{1}
          \and
          A. Caccianiga\inst{2}
          \and
          P. Severgnini\inst{2}
          \and
          R. Della Ceca\inst{2}
          }

   \institute{INAF-Osservatorio Astronomico di Cagliari, Via della Scienza 5, 09047, Selargius (CA), Italy\\
              \email{pcastang@oa-cagliari.inaf.it}
         \and
             INAF-Osservatorio Astronomico di Brera, Via Brera 28, 20121, Milan, Italy\\
             }

   \date{}

 
  \abstract
                   {A relationship between the water maser detection rate and large nuclear column densities in AGN has often been cited in the literature. Indeed, detailed studies of luminous water masers, typically associated with the nuclear activity, allow us to investigate the innermost regions of AGN, with an impact on the still debated Unified Model for this class of objects.}
   {We have recently entertained a search for maser emission in a well-defined sample of Compton-thick AGN aimed at investigating, on firm statistical bases, the aforementioned relationship. While the survey is still ongoing, and is matter of a forthcoming publication, a new luminous water maser has been detected in the lenticular (field) S0 galaxy IRAS15480-0344, whose origin, associated with an accretion disc or a nuclear outflow/jet, needs to be assessed.}
   {Multi-epoch single-dish observations and VLBI measurements were performed to investigate the distribution, spatial extension, and variability of the maser emission in order to infer the main characteristics of the water megamaser.}
   {The new detection in IRAS15480-0344 is reported: a megamaser with a total single-dish isotropic luminosity of $\sim$200\,L$_{\odot}$ and a profile composed of two main features, a broad line with a full width to half maximum (FWHM) linewidth of $\sim$90\,km\,s$^{-1}$ and a narrow (FWHM$<$1\,km\,s$^{-1}$) one. We performed a follow-up to the detection with the Very Long Baseline Array (VLBA) and confidently detected only the narrow component, which is coincident with the nuclear radio continuum emission detected with the Very Large Array at 8.4\,GHz. A weak narrow feature has also been detected in the velocity range of the broad feature and is located 15\,pc to the north-west with respect to the stronger component. Neither maser spot is associated with the compact radio continuum sources derived from the same VLBA dataset.}
   {The different line profiles and the spatial separation between the two features in the spectrum of IRAS15480-0344 suggests a composite origin for the maser. The large linewidth and the strong spectral variability indicate that the broad component might originate from the interaction between a radio jet and ambient molecular clouds. Based on its small linewidth and the absence of high velocity features, instead, we favour an outflow origin for the narrow component. This scenario is consistent with the hypothesis of the presence of strong nuclear outflows recently invoked to explain the main characteristics of field S0 galaxies.} 


   \maketitle
%

\section{Introduction}

The Unified Model of active galactic nuclei (AGN) is based on the premise the all AGN are intrinsically the same object and that the observed dichotomy between broad line (``type~1'') and narrow line (``type~2'') AGN is only due to the orientation relative to a dusty toroidal structure surrounding the nuclear engine (the ``torus'') \citep{antonucci93,urry95}. According to this scheme, objects viewed face-on in which the line of sight to the nucleus is clear are classified as type~1 AGN, while those observed edge-on are recognized as type~2 because the torus obscures the broad line region (BLR) that becomes visible only in polarized light \citep[e.g. NGC~1068; ][]{antonucci85}. Although the existence of an axisymmetric absorber and its crucial role in determining the observed diversity among AGN has been widely confirmed \citep[e.g.][]{jaffe04,tristram07}, its composition and geometrical structure, as well as its dynamical origin, are not yet fully understood. Indeed, some observational evidence calls for a revision of the standard, uniform torus, seen as a hydrostatic, separate entity. Column density ($N_{\rm H}$) variability measurements on a sample of obscured AGN revealed that the X-ray absorber must be clumpy and located at sub-parsec distance from the black hole \citep{risaliti02}, i.e. inside the dust sublimation radius. The physical parameters derived for one of the best studied objects (i.e. NGC~1365) suggest that the X-ray absorber and the BLR clouds are one and the same \citep{risaliti09}. Infrared (IR) properties of AGN provide an independent clue on the clumpiness of the obscuring medium. In fact, recent models of AGN dusty tori demonstrate that a clumpy structure can better account for the IR spectral energy distribution (SED) of AGN (requiring dust at multiple temperatures) and the compact size inferred from mid-IR interferometry \citep[e.g.][]{hoenig06,nenkova08}. In addition, X-ray and infrared reverberation observations suggest that the torus is only the outer part of a continuous distribution of clouds, whose properties reflect their changes in composition across the dust sublimation radius ($R_{\rm d}$): from atomic and ionized at $r < R_{\rm d}$, producing the broad emission lines, to molecular and dusty at $r > R_{\rm d}$, obscuring the optical/ultraviolet radiation and re-emitting it in the IR \citep[][and references therein]{elitzur08}. Such a structure is implicit in disc-wind models, which foresee the presence of an outflowing population of clouds embedded in a hydromagnetic wind originating from the accretion disc surface \citep[e.g.][]{koenigl94}. Although a detailed dynamical model of this clumpy outflow is not yet available, a natural consequence of the ``disc-wind scenario'' \citep[][and references therein]{elitzur08} is the disappearance of the torus and the BLR at low accretion rates, which might explain the existence of ``true'' Seyfert~2, i.e. object that intrinsically lack a BLR (they are unobscured in the X-rays and do not show any broad line in polarized light). The disc-wind paradigm also provides a natural explanation for the trend exhibited by the intrinsic broad line emission to follow an evolutionary sequence as the accretion rate decreases \citep{elitzur2014}. Moreover, magnetically driven winds are considered the best candidates for the origin of the highly ionized, high density, fast outflowing winds found in bright AGN \citep[][; and references therein]{slone2012}. 

The knowledge of physical properties, the structure, and the kinematics of the matter in the vicinity of supermassive black holes (SMBH) is essential to build detailed models of the clumpy outflow and to test the disc-wind scenario. While X-ray variability studies can provide accurate information on the atomic and ionized matter on scales of the BLR, the radio emission from luminous H$_2$O masers (the so-called ``megamasers'') constitutes a fundamental instrument to study the geometry and kinematics of the molecular gas at sub-parsec distance from SMBH. H$_2$O masers may trace distinct regions in the AGN environment, from nearly edge-on accretion discs to nuclear outflows in the form of jets or winds \citep[for recent reviews see][]{henkel05,lo05, greenhill07,tarchi2012}. Very Long Baseline Interferometry (VLBI) and single-dish monitoring studies of disc-masers allow us to map accretion discs and to determine the enclosed dynamical masses \citep[e.g.][]{kuo2011}. Jet-masers observations, instead, can provide estimates of the velocity and density of jet material \citep{peck03}. H$_2$O maser emission have been also found to be associated with nuclear winds at < 1pc from the nuclear engine. In particular, water maser observations in Circinus \citep{greenhill03} and NGC~3079 \citep{kondratko05} seem to have resolved individual outflowing torus clouds. In fact, \citet{greenhill03} discovered that the H$_2$O masers in Circinus trace both a Keplerian disc and a wide-angle outflow which appears to be collimated by the warps of the disc. In NGC~3079, VLBI observations of the maser emission revealed a clumpy thick disc. In addition to that, four maser features were found to be located at high latitude above the disc (at $\sim$0.5\,pc from the disc plane) and interpreted as part of a nuclear wind by \citet{kondratko05}. Proper motion measurements and comparison of these outflow-masers with their disc counterpart provide the most promising method for probing the structure and kinematics of the torus molecular clouds \citep{nenkova08}.

As part of a search for water maser emission in a well defined sample of Compton-thick (CT) AGN \citep{severgnini2012}, we detected a luminous H$_2$O maser in the mid-IR-bright Seyfert~2 galaxy IRAS~15480-0344 (hereafter IRAS15480). The results of the entire survey and their statistical analysis will be reported in a forthcoming paper (Paper~II., in prep.). Here we present the results of 22\,GHz single-dish and VLBI observations of the newly found water maser in IRAS15480. Section~\ref{sect:IRAS15480} summarize the main properties of IRAS15480. Section~\ref{sect:obs} describe the details of the single-dish and VLBI observations and data reduction. The results are reported in Section~\ref{sect:results}. Finally, in Section~\ref{sect:discussion} and~\ref{sect:conclusions} we discuss the possible origin of the water maser emission in IRAS15480 and draw our conclusions. Throughout the paper we adopt a cosmology with $\Omega_{\rm M} =0.3$, $\Omega_{\rm \Lambda} =0.7$ and $H_0 = 70$\,km\,s$^{-1}$\,Mpc$^{-1}$.   

\section{The Compton-thick Seyfert galaxy IRAS15480}\label{sect:IRAS15480}
IRAS15480 is a lenticular \citep[S0;][]{malkan98} galaxy located at a distance of $\sim$130\,Mpc\footnote{Hubble flow distance calculated using the optical redshift $z$=0.0303 \citep{degrijp92}.}. It harbors a Seyfert~2 nucleus with weak evidence for broad H$\alpha$ in polarized light \citep{young96}. A broad H$\alpha$ component has also been tentatively detected in total flux \citep{young96}. As mentioned before, IRAS15480 is a Compton-thick object, with an $N_{\rm H} > 1 \times 10^{24}$\,cm$^{-2}$ \citep{severgnini2012,brightman2011,guainazzi05}. The detection of a silicate absorption feature in the Spitzer IR spectrum of IRAS15480 is consistent with the presence of an obscuring torus viewed nearly edge-on \citep{gallimore2010}. The weakness of such feature is further compatible with clumpy torus models, as the one presented in \citep{nenkova08}. Emission from hot dust, leaking out through a clumpy obscuring structure, might also explain the NIR excess present in the nuclear SED of IRAS15480 \citep{videla2013}. Molecular gas has been detected towards IRAS15480 in the CO $J=2 \rightarrow 1$ transition, with centroid velocity 9039\,km\,s$^{-1}$ \citep{strong04}. At radio frequencies, the nuclear emission is unresolved in the highest resolution radio maps available, so far, taken with the Very Large Array (VLA) at 8.4 GHz. The nuclear radio source has an angular diameter $<$0.1\arcsec, corresponding to $<$60\,pc \citep{thean00,schmitt01}. No VLBI radio continuum maps are reported, so far, in literature.

\section{Observations and data reduction}\label{sect:obs}

\subsection{Single-dish observations}

Observations of the 6$_{16}$--5$_{23}$ transition of ortho-H$_2$O (rest frequency 22.23508\,GHz) towards IRAS15480, were conducted with the Robert~C.~Byrd Green Bank Telescope (GBT) of the National Radio Astronomy Observatory (NRAO), on April 7, April 12, and May 12, 2012. We used two of the seven beams of the K-band focal plane array (KFPA) receiver in total power nod mode, keeping one of the two beams alternatively on source during integration. The KFPA receiver provides dual circular polarization. We configured the spectrometer with two 200\,MHz IFs, each with 8192 channels, yielding a channel spacing of 24\,kHz, corresponding to $\sim$0.3\,km\,s$^{-1}$ at the frequency of 22\,GHz. The first spectral window was centered at the frequency corresponding to the recessional velocity of the galaxy and the second was offset by 180\,MHz to the red. This set up gives a total frequency coverage of 380\,MHz ($\sim$5100 \,km\,s$^{-1}$ at 22\,GHz). The full width to half power (FWHP) beamwidth was $\sim$30\arcsec and the pointing accuracy was, on average, 5\arcsec. 

The data were reduced and analysed using GBTIDL\footnote{http://gbtidl.nrao.edu/}. In order to improve the signal-to-noise ratio, we smoothed the blank sky reference spectrum with a 16 channel boxcar function, before the calibration. This procedure has the advantage to enhance the signal-to-noise ratio without degrading the spectral resolution. We then calibrated the spectra utilizing standard routines and applying the default zenith opacity and gain curve. The uncertainty in this flux calibration procedure is estimated to be no more than 20\%. We averaged the individual scans and the two polarizations to produce the final spectrum. We then subtracted a polynomial baseline, typically of degree between 3 to 6, from the spectra.

\subsection{VLBI observations, archival data, and imaging}\label{sect:vlbaobs}

The Very Long Baseline Array (VLBA) of the NRAO, observed IRAS15480 on June 2 and 3, 2012, in two tracks of 5 hours each (project code BC207). The data were taken with two 16\,MHz IFs in dual circular polarization. The two IFs were centered at the velocity of the two maser features (Table~\ref{table:i15480} and Fig.~\ref{fig:i15480}). Cross-correlation of the data was performed using the DiFX software correlator with 128 channels per IF and polarization, yielding a channel spacing of 125\,kHz (corresponding to $\sim$1.7\,km\,s$^{-1}$ at 22\,GHz). A second DiFX pass in ``spectral zooming'' mode provided a channel spacing of 15.6\,kHz ($\sim$0.2\,km\,s$^{-1}$) over a 2\,MHz band centered on the velocity of the narrow maser component. We observed in phase-referencing mode, to correct phase variation caused by the atmosphere. We used J1555-0326 as a phase calibrator. We also observed the strong compact quasars J0927+3902 and J1800+3848 at intervals of 1hr and 20 minutes in each observing session, in order to correct instrumental single-band delays. We have also examined archival VLBA continuum data at 4.98\,GHz (program code BS104). These observations were taken in January 2002, in phase-referencing mode (phase reference source J1555-0326). Data were recorded at 128 Mbps, using 4 IFs of 8 MHz, and 16 channels (2 bits/sample). The strong compact sources J1557-0001 and J1642+3948, were observed as fringe finders.

\begin{table*}
\caption{VLBA maps of IRAS15480.}
\label{table:maps}      
\centering          
\begin{tabular}{crrlcrc}     
\hline\hline
Band & $\nu$ & Channel width  & Weights & Synthetized beam & P.A.         & r.m.s \\ 
     & (GHz) & (km\,s$^{-1}$)  &         & HPBW (mas)       & (\degr)      & (mJy/beam) \\ 
\hline      
K    & 21.571 & 1.8           & natural & 1.8$\times$1.0    & -10          & 2.2 \\
     & 21.581 & 1.8           & natural & 1.8$\times$1.1    & -7           & 2.1 \\
     & 21.579 & 0.2           & natural & 1.8$\times$1.1    & -7           & 5.1 \\
     & 21.584 & ...           & natural & 1.8$\times$1.0    & -10          & 0.1 \\
     &        & ...           & tapered\tablefootmark{a} & 2.2$\times$1.5    & 9            & 0.1 \\
C    & 4.983  & ...           & natural & 4.5$\times$1.6    & -14          & 0.7 \\
\hline
\end{tabular}
\tablefoot{
\tablefoottext{a}{Taper 200\,M$\lambda$.}
}
\end{table*}

Data reduction and analysis were carried out utilizing the NRAO Astronomical Image Processing System (AIPS\footnote{http://www.aips.nrao.edu/}). The first step of the data reduction was to apply the latest value of the Earth's orientation parameters. Then, we corrected the sampler voltage offsets and used the measured antenna gains and system temperatures to calibrate the amplitudes. After correcting the interferometer delays and phases for the effect of diurnal feed rotation (parallactic angle), we removed the instrumental delays caused by the passage of the signal through the electronics. The latter step was achieved fitting the fringe patterns from J0927+3902 and J1800+3848 to compute delay and phase residuals for each antenna and IF. At this point, we calibrated the bandpass shape, using J0927+3902 as a bandpass calibrator, and corrected the frequency axes of the maser source to compensate for the Doppler shift caused by the antennas motion with respect to the sun. Finally, we fringe fitted the data from the phase reference source J1555-0326 to solve for the atmospheric phase variation. The rate solutions were then interpolated and applied to our target source IRAS15480. We calibrated each track and spectral mode separately, with the exception of the final phase calibration of the data in ``spectral zooming'' mode, where we used the solution of the fringe fitting on the reference source calculated from the lower resolution dataset. Then, the calibrated data from the two tracks were combined for imaging. The data were Fourier-transformed using natural weighting and deconvolved using the CLEAN algorithm \citep{hoegbom74}. We mapped a field of 0.8$\times$0.8 arcseconds$^2$ (corresponding to $\sim$480$\times$480\,pc$^2$) centered at the position of the optical nucleus of IRAS15480. We obtained three images from the line dataset, one for each of the two IFs in the standard mode and one for the ``spectral zooming'' mode. A continuum image was obtained averaging the line-free channels of both IFs. Table~\ref{table:maps} reports, for each map, the observing band, the central frequency, and the channel width. The weights used to produce the maps are also indicated, together with the dimensions and position angles of the synthetized beams, and the average r.m.s per channel.  

For the continuum experiment, the calibration was performed following the same steps, except that pulse-cal information were recorded and used to correct for instrumental delays. Bandpass and Doppler shift corrections were not necessary in the continuum case. The data were imaged using natural weighting.

\section{Results}\label{sect:results}

\subsection{Detection and variability of the water maser}

\begin{figure}
\centering
\includegraphics[width=\hsize]{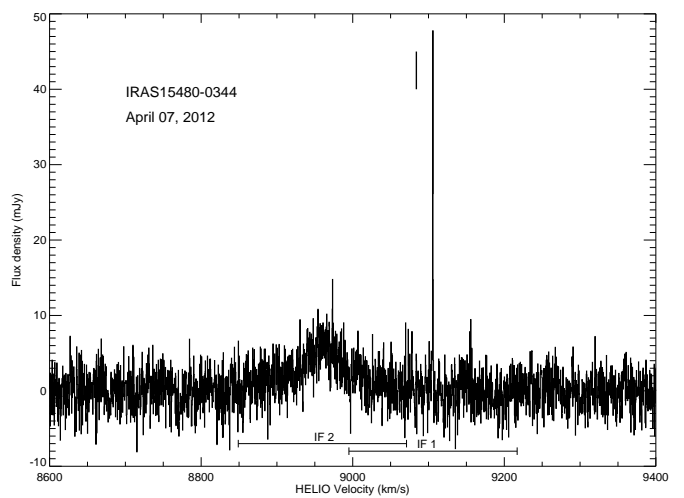}
\caption{H$_2$O maser profile observed towards IRAS\,15480-0344 with the GBT on April 7, 2012. Channel spacing is 0.3\,km\,s$^{-1}$. The vertical line marks the recessional velocity of the galaxy. The horizontal lines indicates the velocity ranges covered by the two IFs in the VLBA observation. The velocity scale is with respect to the heliocentric frame and uses the optical convention.}
\label{fig:i15480}
\end{figure}

\begin{figure}
\centering
\includegraphics[width=\hsize]{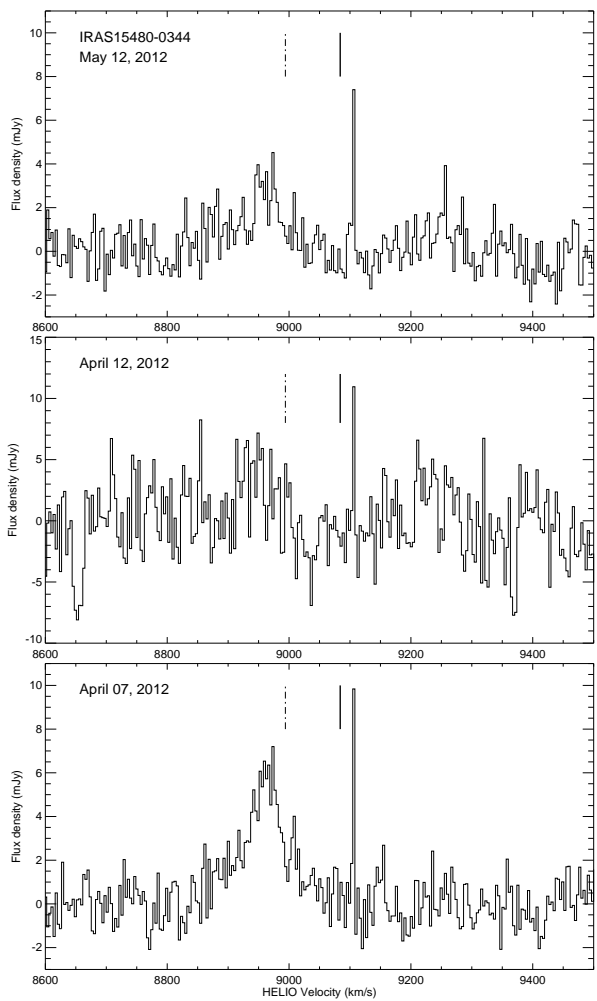}
\caption{H$_2$O maser spectra observed towards IRAS\,15480-0344 with the GBT on April 7 {\it (bottom panel)}, April 12 {\it (middle panel)}, and May 12, 2012 {\it(top panel)}. Channel spacing is 3.3\,km\,s$^{-1}$. The vertical line indicates the recessional velocity of the galaxy. The dash-dotted line marks the velocity of the maser spot M2, detected with the VLBA (Sect.~\ref{sect:res_vlba}). The velocity scale is with respect to the heliocentric frame and uses the optical convention.}
\label{fig:i15480_comp}
\end{figure}

On April 7, 2012, we detected water maser emission in IRAS15480. The maser spectrum shows two main features (Fig~\ref{fig:i15480} and Table~\ref{table:i15480}): a broad component with a full width at half maximum (FWHM) linewidth of $\sim$90\,km\,s$^{-1}$, blueshifted by $\sim$120\,km\,s$^{-1}$ with respect to the recessional velocity of the galaxy, V$_{sys} =cz=9084$\,km\,s$^{-1}$, and a narrow (FWHM$<$1\,km\,s$^{-1}$) line at a velocity of $\sim$9106\,km\,s$^{-1}$, thus very close to the systemic velocity. The total isotropic luminosity is $\sim$200\,L$_{\odot}$. The narrow component was detected also in the observing sessions of April 12 and May 12, 2012. Line properties (velocity, linewidth and peak flux density) remained constant within the errors in the three epochs (Table~\ref{table:i15480}). The broad emission feature is still visible in the spectrum taken in May (Fig~\ref{fig:i15480_comp}) but appears to be significantly weaker. The line peak and integrated flux density decreased to half of their initial values in approximately one month, indicating strong variability. 

Water maser emission was already searched, unsuccessfully, in this galaxy with the GBT in 2003 (Braatz et al. 2004) and with the 70-m at Goldstone in 1997 (Greenhill et al. 1997) with r.m.s of 5 and 20 mJy for a 0.3\,km\,s$^{-1}$ channel, respectively. While the latter observation has not detected the maser line(s) likely due to the coarse sensitivity of the spectrum, the former survey should have clearly seen, at least, the bright narrow feature. This result leads us to conclude that not only the broad component but also the narrow emission feature is variable.

\begin{table*}
\caption{Line parameters of the H$_2$O maser in IRAS15480}             
\label{table:i15480}      
\centering          
\begin{tabular}{c c c c c l l l l}     
\hline\hline       
Obs. date & Telescope & $rms$ & Channel width & $V_{\rm HEL}$     & $FWHM$        & $S_{\rm peak}$ & $\int S dV$       & $L_{\rm H2O}$ \\
          &           & (mJy) & (km\,s$^{-1}$) & (km\,s$^{-1}$)   & (km\,s$^{-1}$) &  (mJy)       & (mJy km\,s$^{-1}$) & (L$_{\odot}$) \\
\hline
2012-Apr-07 & GBT      & 3    &  0.3          & 9105.75$\pm$0.02 & 0.63$\pm$0.04 & 48$\pm$3     & 35$\pm$4          & 14            \\
            &          & 2    &  1.2          & 8960$\pm$2       & 87$\pm$5      & 5.1$\pm$0.2  & 509$\pm$49        & 197           \\
2012-Apr-12 & GBT      & 7    &  0.3          & 9105.73$\pm$0.04 & 0.55$\pm$0.08 & 57$\pm$7     & 37$\pm$10         & 14           \\
2012-May-12 & GBT      & 2    &  0.3          & 9105.75$\pm$0.02 & 0.58$\pm$0.03 & 43$\pm$2     & 26$\pm$3          & 10            \\
            &          & 1    &  1.8          & 8958$\pm$4       & 74$\pm$10     & 2.4$\pm$0.3  & 219$\pm$57        & 85            \\
2012-Jun-06 & VLBA     & 5    &  0.2          & 9105.88$\pm$0.02 & 0.46$\pm$0.02 & 47$\pm$4     & 23$\pm$3          &  9            \\
            &          & 2    &  1.8          & 8994$\pm$2       & $\sim$2       & 12$\pm$2     & $\sim$24          &  9             \\
\hline                  
\end{tabular}
\end{table*}

\subsection{VLBA results}\label{sect:res_vlba}

\begin{figure*}
\centering
\includegraphics[width=17cm]{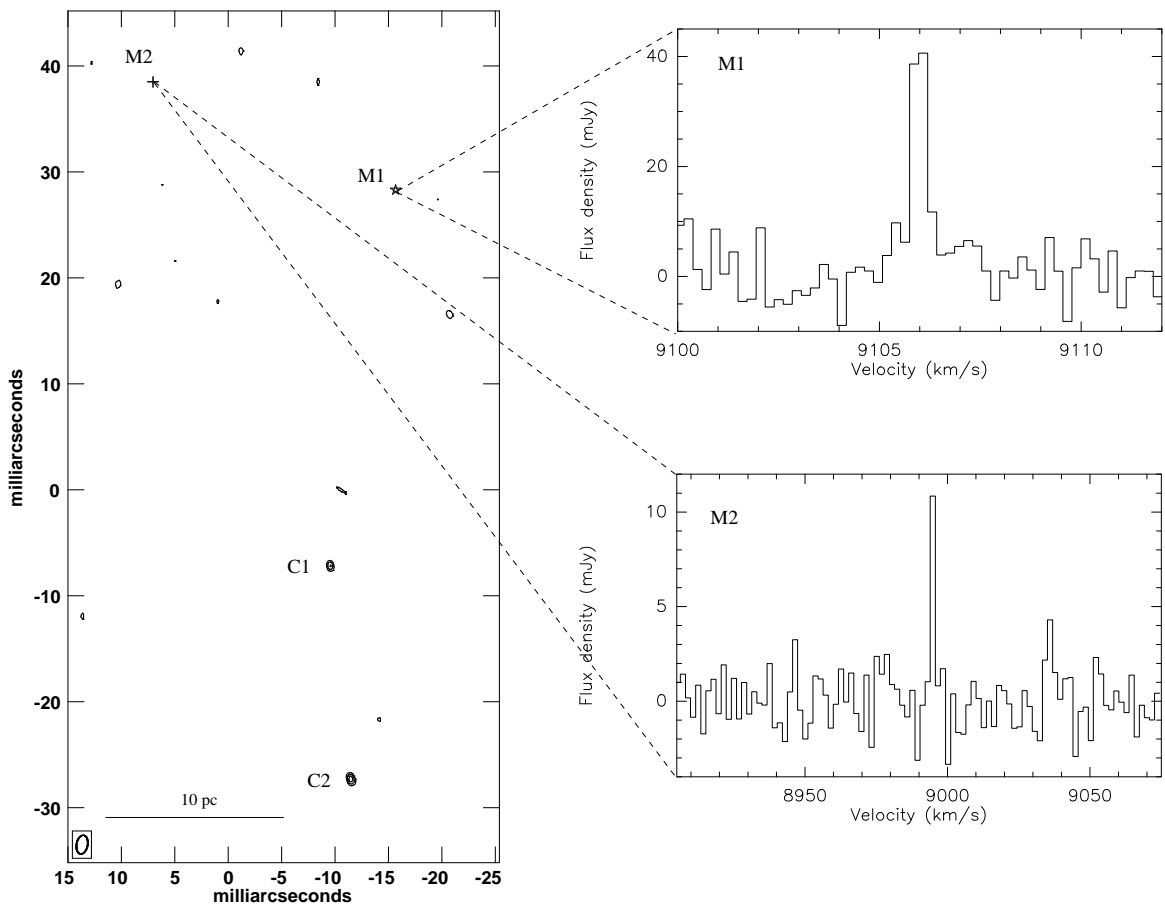}
\caption{{\it Left panel:} 22\,GHz VLBA contour map of the radio continuum emission from the nucleus of IRAS15480. Contour levels are 4, 5, 6, 7 $\times$ 0.1\.mJy/beam. The star indicates the position of the brightest maser spot (M1) at $v=9106$\,km\,s$^{-1}$, while the plus sign marks the position of the weaker maser feature (M2) with a velocity of 8994\,km\,s$^{-1}$. The positions are plotted relative to the centre of the map (i.e. the position of the optical nucleus of IRAS15480). {\it Top right panel:} VLBA spectrum of the brightest maser spot M1. The channel spacing is 0.2\,km\,s$^{-1}$. {\it Bottom right panel:} VLBA spectrum of the blueshifted maser feature M2. The channel spacing is 1.8\,km\,s$^{-1}$.}
\label{fig:i15480_vlba}
\end{figure*}

The narrow line feature was successfully detected with the VLBA with a signal-to-noise ratio (S/N) of about 10, both in the low resolution dataset (channel spacing $\sim$1.8\,km\,s$^{-1}$) and in the dataset in ``spectral zooming'' mode (channel spacing $\sim$0.2\,km\,s$^{-1}$). In the latter, the maser emission is slightly resolved in velocity and has a peak flux density of about 45\,mJy  (Fig.~\ref{fig:i15480_vlba}; upper right panel). The velocity-integrated map shows that the line emission arises from a single spatially unresolved spot, located at $\alpha_{2000}$=15$^{\rm h}$50$^{\rm m}$41$^{\rm s}$.49695 and $\delta_{2000}$=$-$03\degr53\arcmin18\arcsec.0217 (M1, Fig.~\ref{fig:i15480_vlba}; left panel). The uncertainty in the absolute position is of 0.2 and 0.4 mas in right ascension and declination, respectively, and is dominated by the uncertainty in the position of the reference source, J1555-0326 (Sect.~\ref{sect:vlbaobs}). The position of M1 is coincident, within the relative position error of 30\,mas, with the position of the nuclear radio continuum source visible in the VLA X-band image \citep{schmitt01}. In the frequency band centered at the velocity of the broad emission feature (8850--9070\,km\,s$^{-1}$) we detected only one maser spot, with a peak flux density of $\sim$12\,mJy, at $V_{\rm HEL}$=8994\,km\,s$^{-1}$ (Fig.~\ref{fig:i15480_vlba}; lower right panel). The line emission is unresolved both in space and velocity and is located at a distance of $\sim$25\,mas ($\sim$15\,pc) to the north-west of the strongest maser feature (M2, Fig.~\ref{fig:i15480_vlba}; left panel). No other maser emission was detected in the low resolution image cubes, in the total inspected velocity range (8850--9220\,km\,s$^{-1}$), above a 5\,$\sigma$ noise level of 10\,mJy/beam. A few ``tentative'' spots, with peak flux densities of $\sim$10\,mJy/beam, have been detected at distances larger than 100\,mas (or 60\,pc) from the nuclear position and are likely to be noise spikes, therefore, they will not be considered in the following. 

\begin{table*}
\caption{Parameters of the nuclear maser sources detected in the VLBA 22\,GHz map of IRAS15480.}             
\label{table:vlba_maser}      
\centering          
\begin{tabular}{c c c r c c c}     
\hline\hline       
Label         & RA                      & Dec.                & Size               & $V_{\rm HEL}$    & $S_{\rm peak}$ & $T_{\rm B}$        \\
              & 15$^{\rm h}$ 50$^{\rm m}$ & -03\degr 53\arcmin  & (mas $\times$ mas) & (km\,s$^{-1}$)  & (mJy/beam)    & (K)               \\
\hline
M1            & 41$^{\rm s}$.49695       & 18\arcsec.0217     & 0.9$\times$0.5     & 9015.99$\pm$0.02 & 48$\pm$6     & 6.8$\times$10$^7$ \\
M2            & 41$^{\rm s}$.49847       & 18\arcsec.0115     & 0.9$\times$0.5     & 8994$\pm$2       & 12$\pm$2     & 1.7$\times$10$^7$ \\
\hline                  
\end{tabular}
\end{table*}

\begin{figure*}
\centering
\includegraphics[width=17cm]{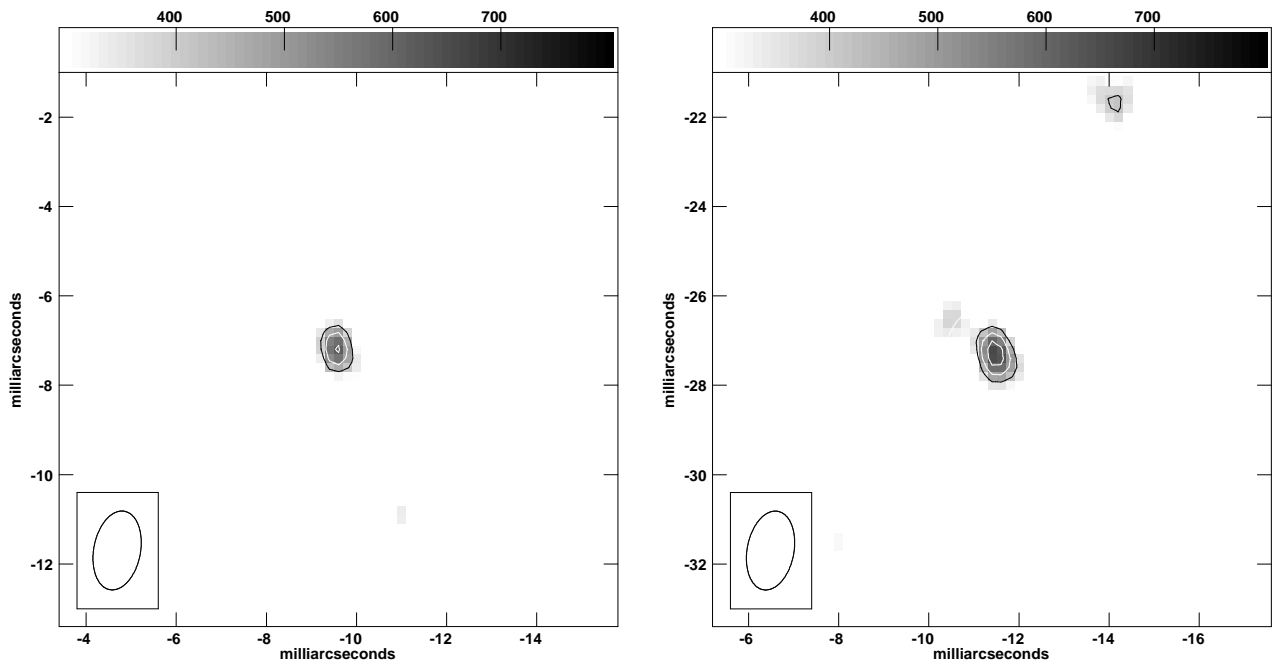}
\caption{22\,GHz VLBA greyscale plus contour map, obtained using natural weighting, of the continuum sources C1 ({\it left panel}) and C2 ({\it right panel}). Greyscale flux range is 300--800\,$\mu$Jy. Contour levels are 4, 5, 6, 7 $\times$ 0.1\.mJy/beam.}
\label{fig:vlba_cont}
\end{figure*}

\begin{figure}
\centering
\includegraphics[width=\hsize]{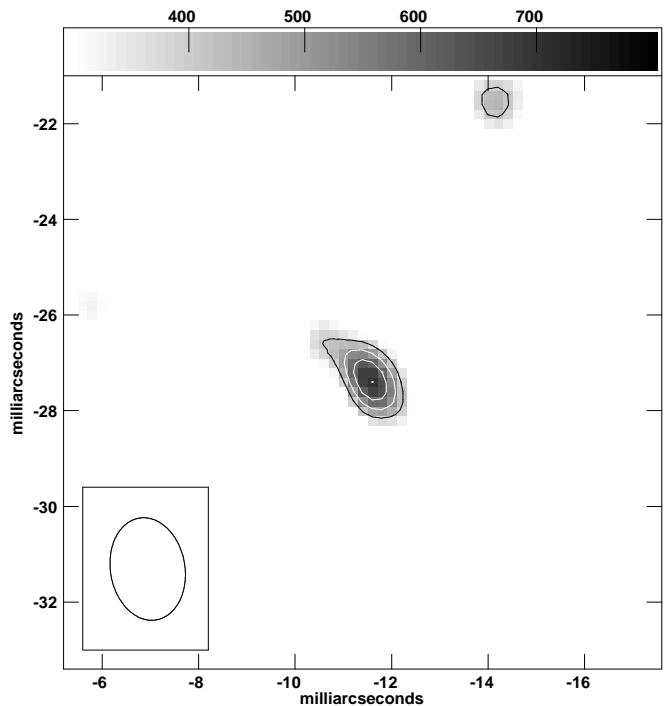}
\caption{22\,GHz VLBA greyscale plus contour map, obtained using an uvtaper of 200\,M$\lambda$, of the continuum sources C2. Greyscale flux range is 300--800\,$\mu$Jy. Contour levels are 4, 5, 6, 7 $\times$ 0.1\.mJy/beam.}
\label{fig:vlba_c2}
\end{figure}

A radio continuum K-band image was also produced averaging the line-free channels, using natural weighting, reaching a r.m.s of 0.1\,mJy/beam (Table~\ref{table:maps}). Within a radius of of $\sim$50\,mas from the reference position (a region roughly coincident with the dimension of the VLA source) we detected 2 radio continuum sources labelled ``C1'' and ``C2'' (Fig.~\ref{fig:i15480_vlba}; right panel). A two-dimensional Gaussian fit to the individual components reveals that they are both unresolved. However, the inspection of the greyscale map of source C2 suggests the presence of weak emission elongated towards the north-west (Fig~\ref{fig:vlba_cont}). In order to recover part of the diffuse emission, we have also produced a tapered map (uvtaper 200\,M$\lambda$) that clearly shows that component C2 is slightly resolved along the major axis (Fig~\ref{fig:vlba_c2}), with dimensions 1.8$\times$0.8\,mas (Table~\ref{table:vlba_cont}). The peak flux densities of components C1 and C2 are 0.6 and 0.7\,mJy\,beam$^{-1}$, respectively, implying a radio power at 22\,GHz of about 1.4$\times$10$^{21}$\,W\,Hz$^{-1}$ and brightness temperatures larger than 3.6$\times$10$^6$\,K. We have also reduced archival C-band VLBA data but no continuum source has been detected above a 5$\sigma$ noise level of 3.5\,mJy/beam. 

\begin{table*}
\caption{Parameters of the nuclear continuum sources detected in the VLBA 22\,GHz map of IRAS15480.}             
\label{table:vlba_cont}      
\centering          
\begin{tabular}{c c c r c c c}     
\hline\hline       
Label         & RA                      & Dec.                & Size               & $S_{\rm peak}$ & $T_{\rm B}$        & log $P$ \\
              & 15$^{\rm h}$ 50$^{\rm m}$ & -03\degr 53\arcmin  & (mas $\times$ mas) & (mJy/beam)    & (K)               & (W\,Hz${-1}$) \\
\hline
C1            & 41$^{\rm s}$.49736       & 18\arcsec.0572     & 0.9$\times$0.5     & 0.6$\pm$0.1   & 3.6$\times$10$^6$ & 1.4$\times$10$^{21}$ \\
C2            & 41$^{\rm s}$.49723       & 18\arcsec.0773     & 1.8$\times$0.8     & 0.7$\pm$0.1   & 3.6$\times$10$^6$ & 1.4$\times$10$^{21}$  \\
\hline                  
\end{tabular}
\end{table*}

\section{Discussion}\label{sect:discussion}

\subsection{IRAS15480 nuclear radio continuum emission}\label{sect:cont_discu}
Radio continuum emission was detected towards the nucleus of IRAS15480 with the VLA at 1.4 and 8.4\,GHz. The nuclear source is unresolved at both frequencies and the reported flux densities, $S_{1.4}$=42\,mJy \citep[NVSS;][]{condon98} and $S_{8.4}$=11\,mJy \citep{schmitt01}, imply a spectral index $\alpha_{8.4}^{1.4}$=0.7 ($S \propto \nu^{-\alpha}$), typical of optically thin synchrotron emission. Using this value of the spectral index, we obtain a 4.9\,GHz flux density of 17\,mJy and a radio luminosity $L_{4.9}$=1.7$\times$10$^{39}$\,erg\,s$^{-1}$. We then employ the ratio $R_{\rm X}= L_{4.9 \rm GHz}/L_{2-10 \rm keV}$, where $L_{2-10 \rm keV}$ is the unabsorbed X-ray luminosity in the 2--10\,keV band, as a measure of the radio loudness \citep{terashima03}. The intrinsic X-ray luminosity of IRAS15480 has been estimated to be 1--4\,$\times$10$^{43}$\,erg\,s$^{-1}$ \citep{brightman2011} that yields $\log R_{\rm X} \sim -4$, slightly above the approximate boundary between radio-loud and radio-quiet AGN set by \citet{terashima03}. 

We note, however, that this boundary was derived considering the nuclear radio emission that, for this particular case, at VLA resolution is contaminated by extended emission as indicated by the significant difference between the flux densities at VLA and VLBI scales (see next paragraph). Therefore, the value of $\log R_{\rm X}$ we obtain should be considered an upper limit. Using the upper limit to the VLBA flux at 4.9\,GHz (Sect.~\ref{sect:res_vlba}), instead of the VLA flux, gives $\log R_{\rm X} \sim -4.5$. This suggests that IRAS15480 is either a radio quiet AGN or an object with a radio luminosity intermediate between radio-loud and radio-quiet AGN. While the origin of the radio emission in radio loud AGN is well understood and is believed to be synchrotron radiation emitted  by relativistic particles in jets and lobes, VLBI observations revealed that different physical mechanisms can be at work in radio quiet AGN. The properties of radio emission suggest a number of radio quiet objects might be scaled-down version of more luminous AGN \citep[e.g.][]{caccianiga01,giroletti09}. However, in  NGC~1068, for example, thermal emission from an X-ray ionized corona or wind arising from the disc (in addition to the jet emission) has been invoked to explain part of the radio emission from the nucleus \citep{gallimore04}. 

In our VLBA continuum image at 22\,GHz of the nuclear region of IRAS15480, we detected two compact continuum sources (``C1'' and ``C2''; Fig.~\ref{fig:i15480_vlba}), coincident, within the errors, with the VLA source at 8.4\,GHz. The total VLBI flux density at 22\,GHz is about 23\% of the expected VLA flux density at same frequency, calculated assuming a constant spectral index of 0.7 in the interval 1.4--22\,GHz. Furthermore, from our analysis of archival VLBA data, we derived an upper limit of 3.5\,mJy for the continuum flux density at 4.9\,GHz, which implies a VLA over VLBI flux density ratio $\la$0.2. This indicates that most of the radio continuum emission in IRAS15480 is resolved at VLBI scales, a phenomenon that is common in Seyfert galaxies \citep[e.g.][]{panessa2013}. The brightness temperatures of ``C1'' and ``C2'' is in excess of 3.6$\times$10$^6$\,K and tend in favour of a non-thermal origin for these sources. From their non-detection at 4.9\,GHz, we infer a spectral index $\alpha_{22.2}^{4.9} < 1$, consistent with emission from outflow or jet-base features found in other Seyfert galaxies \citep[e.g.][]{giroletti09}. An origin from radio supernovae (RSNe) or supernova renmants (SNRs), although in agreement with the brightness temperature and spectral index limits, is unlikely due to the power of the radio continuum sources in IRAS15480 ($P_{22}$=1.4$\times$10$^{21}$\,W\,Hz$^{-1}$). Indeed, the most luminous RSNe and SNRs found, so far, in external galaxies have a 5\,GHz radio power of less than 10$^{21}$\,W\,Hz$^{-1}$ \citep[][and references therein]{castangia08} and are even weaker at higher frequencies, if we assume a classical spectral index of 0.7. Only the extremely bright SN~1986~J, in NGC~891, reached a 22\,GHz luminosity comparable to that of ``C1'' and ``C2'' \citep[$P_{22} \leq 7.0 \times 10 ^{20}$\,W\,Hz$^{-1}$, using the flux densities reported in][]{weiler90}. In order for the compact continuum sources in IRAS154580 to be interpreted as RSNe or young SNRs, we should assume an inverted spectral index and/or that they belong to the class of extremely luminous radio supernovae of which SN~1986~J is a prototype. Therefore, althoug it cannot be a priori excluded, we consider this possibility improbable.
The radio power at 22\,GHz of the compact sources in the nuclear region of IRAS15480 is also two order of magnitude larger than the average power at 5\,GHz of the VLBI sources detected in a sample of local Seyfert galaxies \citep{panessa2013} and is comparable to that of the brightest Seyferts \citep[e.g. IC\,5063; ][]{morganti07}. 

\subsection{Origin of the maser in IRAS15480}\label{sect:maser_discu}
Our single-dish observations of the maser in IRAS15480 highlight the presence of two emission features with very different properties, a broad, strongly variable, blueshifted line and a very narrow component close to the systemic velocity of the galaxy, which remained nearly constant during our monitoring period.
Broad maser lines displaced by $\sim$100\,km\,s$^{-1}$ from the systemic velocity, such as the one observed in IRAS15480, have been associated with the interaction of the AGN jet with molecular clouds of the parent galaxy \citep{gallimore01,peck03} or have been interpreted as the result of the amplification of the radio continuum emission from the jet by foreground clouds in the AGN torus \citep{sawada08}. In the former case, a remarkable variability of the maser flux density, correlated with variations in the continuum, has been reported \citep{peck03}. On the contrary, the narrow emission line detected in IRAS15480 close to the systemic velocity, resembles the systemic features typically observed in disc-maser spectra \citep[e.g. NGC~4258;][and references therein]{humphreys2013}. H$_2$O masers tracing wide-angle nuclear outflows also show emission lines with FWHM$\sim$1\,km\,s$^{-1}$ and velocities spread over $\sim \pm$160\,km\,s$^{-1}$ of the systemic one \citep[Circinus;][]{greenhill03} or reflecting those of the most proximate side of the accretion disc \citep[NGC~3079;][]{kondratko05}. The different line profiles and variability, therefore, seem to suggest that the maser emission arises from two physically distinct regions within the galaxy, with the broad blueshifted component most likely associated with the interaction of a jet with the interstellar medium of the host galaxy and the narrow emission line at the systemic velocity with an accretion disc or a nuclear outflow. However, we cannot rule out the possibility that both line complexes are part of the maser emission from a Keplerian accretion disc, with the narrow feature representing the systemic group and the broad component the high-velocity group of lines produced in a slowly rotating disc (rotation velocity $\sim$110\,km\,s$^{-1}$).  

An association with star-formation activity cannot be excluded for the narrow systemic feature, even if the large luminosity, in spite of the narrow linewidth, makes this scenario unlikely. In fact, water maser in Galactic star-forming regions commonly display narrow flaring lines but only the outburst in W49N \citep{liljestroem00,honma04} reached an isotropic luminosity of $\sim$1\,L$_{\odot}$. Narrow extragalactic masers associated with star formation as well, do not exceed 4\,L$_{\odot}$ \citep[][and references therein]{castangia08}. If the star formation scenario is still a viable option for the origin of  the narrow maser feature in IRAS15480, an association with a star-forming region is seemingly ruled out for the broad component, on the basis of the observed variability. For this component, we measured a significant decrease in peak and integrated flux density between the initial detection and the observation made on May 12, with the line profile fundamentally unchanged (see Table~\ref{table:i15480}). Therefore, we can estimate the maximum extent of the masing region based on the distance covered in $\sim$35 days at light-speed, to be 0.025\,pc. This size is too small to explain the large luminosity in term of a collection of W49N-like sources. In fact the small dimension of the masing region would require a space density of OB stars much higher than that so far observed.

High angular resolution observations of the maser in IRAS15480 with the VLBA successfully detected the narrow component close to the systemic velocity of the galaxy. Line properties (velocity, linewidth, peak and integrated flux density) are consistent with those derived from the single-dish spectrum  (Table~\ref{table:i15480} and Fig.~\ref{fig:i15480_vlba}, upper-left panel). The small velocity shift of 0.13\,km\,s$^{-1}$ between the last GBT measurement and the VLBA observation, cannot be due to a secular drift of the systemic lines such as the one typical of disc-maser systems. Indeed, if this was the case, a change in the velocity should have been measured also between the two GBT observations of April 7 and May 12, while the GBT velocity determinations are stable within 0.02\,km\,s$^{-1}$ over 5 weeks. The discrepancy might be caused by the use of two different procedures and softwares to calculate the Doppler correction, at the GBT and at the VLBA. However, we were not able to prove this possibility or to explain the discrepancy in velocity otherwise. Only a weak narrow line has been detected in the frequency band centered on the velocity of the broad blueshifted component. The peak velocity of this line (8994\,km\,s$^{-1}$, Table~\ref{table:i15480} and Fig.~\ref{fig:i15480_vlba}), however, is redshifted by $\sim$20\,km\,s$^{-1}$ from the peak velocity of the broad line in the single-dish spectrum (Fig.~\ref{fig:i15480_comp}). This suggests that the broad maser emission faded below the detection threshold of our VLBI experiment, while we may be witnessing the flare of a new component in the same region. Although unlikely, we cannot exclude the possibility that the new feature M2 arises from a completely different region with a similar velocity field to that giving rise to the broad component. Both maser features are unresolved with dimensions $\la 0.5 \times 0.3$\,pc.

The distance of $\sim$15\,pc between the systemic and blueshifted emission features (M1 and M2; Fig.~\ref{fig:i15480_vlba}), somehow weakens the hypothesis that they are both produced in a Keplerian accretion disc. Indeed, masing disc radii measured so far, are typically $\la$1\,pc \citep[e.g.][]{kuo2011}. A rough upper limit to the inner radius of the masing disc in IRAS15480 can be obtained from the X-ray luminosity. Given the observed correlation between the inner radius of the masing discs and the bolometric luminosity \citep{castangia2013}, for an X-ray luminosity of 1--4\,$\times$10$^{43}$\,erg\,s$^{-1}$ (Sect.~\ref{sect:cont_discu}) we expect a maximum radius of $\sim$4\,pc\footnote{We used the same correction factor of 20 employed by \citet{castangia2013} to calculate the bolometric luminosity of IRAS15480.}. Nevertheless, for a radius of 15\,pc and a rotation velocity of 110\,km\,s$^{-1}$, the binding mass would be $\sim 5 \times 10^7$\,M$_{\sun}$, which is not unreasonable compared to the black hole masses derived from disc-maser studies \citep[e.g.][]{kuo2011}. Therefore, although we still tend in favour of a composite origin of the maser emission in IRAS15480, as suggested by the morphology of the single-dish spectrum, we cannot completely reject the alternative scenario in which the entire maser emission is produced in a edge-on accretion disc.

The optical classification of IRAS15480 as lenticular/S0 (Sect~\ref{sect:IRAS15480}) may support the presence of a nuclear outflow, and hence, the outflow option for the maser origin. Indeed, from an inspection of archival X-ray ROSAT data the absence of significant X-ray emission at the position (and neighboring regions) of the galaxy rule out the association of IRAS15480 with a prominent galaxy cluster. Thus, IRAS15480 belongs to the considerable fraction of S0 galaxies occurring in the field, where the effects of collisions and ram pressure stripping, typically invoked to explain the characteristics of S0 galaxies in clusters (absence of spiral structure, gas poor environments, etc...), are expected to be very small. While the discussion is still open on the cause for the absence of significant amount of gas in field lenticular galaxies, internal processes seem to  be relevant candidates, among which the star formation triggering effect of strong outflows from the active galactic nucleus \citep[e.g.][and references therein]{vandenbergh09}.

Noticeably, despite the aforementioned paucity of gas (although not of dust content) in lenticular/S0 galaxy, the fraction of water maser detection in these galaxies is everything but negligible. According to the sample of 78 galaxies reported in \citet{bennert09} and making use of the  classification from NED, the detection rate in S0 type galaxies is of  $\sim$ 9\% (7/78). Thus, speculatively, we may infer that the  characteristics of S0 galaxies (e.g., nuclear outflows and internally  triggered star formation in field objects and/or the peculiar environments for S0s belonging to clusters) may favor the condition for maser action. Interestingly, it has been proposed that accretion rates close to the Eddington limit and the presence of strong nuclear outflows may play a role also in the high maser detection rate found in another ``peculiar'' class of galaxies, namely the Narrow Line Seyfert~1 (NLSy1), compared to normal Sy1 \citep{tarchi2011}.

\section{Conclusions}\label{sect:conclusions}

During an ongoing search for water maser emission in a well-defined sample of Compton-thick AGN, we detected a new 22\,GHz water megamaser in the lenticular S0 galaxy IRAS15480 at a distance of 130\,Mpc.

We have investigated the luminous megamaser ($L_{\rm H2O}$ $\sim$200L$_{\odot}$) using single-dish spectra taken with the GBT at three epochs and interferometric observations with the VLBA. The single-dish maser spectra display two main maser features, a strong narrow line close to the systemic velocity of the galaxy and a broad line blueshifted from it. Only the narrow feature has been confidently detected in our VLBA maps. The position of this emission, though associated with the nuclear region of the galaxy, is not associated with any obvious centre of activity visible in the 22\,GHz continuum map produced using the line-free channel of the dataset. 

Our interpretation, based on the single-dish profile, the variability of the maser emission, and the accurate position of the maser spots derived from VLBI observations, tends in favor of a jet/outflow origin, consistently with similar cases found in other radio-quiet AGN.The possibility that the whole maser emission is produced in a slowly rotating accretion disc, with a rather large radius (15\,pc), however, cannot be completely rejected. A detailed study of the nuclear environment of IRAS15480 with an angular resolution comparable with that of our VLBA spectral line observations, would be necessary to associate the location of the line emission with that of other sources of activity in the innermost regions of the AGN and, hence, to clarify the physical mechanism at the origin of the maser emission. 

\begin{acknowledgements}
The National Radio Astronomy Observatory is a facility of the National Science Foundation operated under cooperative agreement by Associated Universities, Inc. P.~C. and A.~T. are grateful to A. Sanna, F. Govoni, and G. Surcis for useful discussions. We wish to thank the referee, J. Moran, for a critical reading of the manuscript and useful comments. P.~S. acknowledges financial support from the Italian Ministry of Education, Universities, and Research (PRIN~2010-2011) and from the Italian Space Agency (contract ASI-INAF~I/037/12/0). 
\end{acknowledgements}


\bibliographystyle{aa} 
\bibliography{castangia2016} 

\end{document}